\documentclass[twocolumn,showpacs,superscriptaddress,prb]{revtex4}
\usepackage{amsbsy,amssymb,amsmath,bm}
\usepackage{graphicx,color,ulem}
\usepackage{epsfig}
\usepackage[tight]{subfigure}
\newcommand{\yg}[1]{ \vspace*{0.05in} {\color{blue}\hrule
    \noindent\textsf{Yuri: #1}\hrule}\vspace*{0.05in} }

\newcommand{\curl}{\mathop{\mathrm{curl}}}
\newcommand{\e}{{\text{eff}}}
\newcommand{\fl}{{\rm sat}}
\newcommand{\bl}{\text{bulk}}

\newcommand{\f}[1]{Fig.~\ref{#1}} \newcommand{\eq}[1]{Eq.~(\ref{#1})}
\newcommand{\eqs}[2]{Eqs.~(\ref{#1}) and~(\ref{#2})}

  \def\be{\begin{equation}}
\def\ee{\end{equation}} \def\bea{\begin{eqnarray}}
\def\eea{\end{eqnarray}} \def\l({\left(} \def\r){\right)}

  \renewcommand{\yg}[1]{}

\begin{document}

\title{Flux saturation number of superconducting rings}

\author{D.~V.~Denisov}
\affiliation{Department of Physics and Center for Advanced
Materials and Nanotechnology,    University of Oslo, P. O. Box
1048 Blindern, 0316 Oslo, Norway}
\author{D.~V.~Shantsev}
\affiliation{Department of Physics and Center for Advanced
Materials and Nanotechnology,    University of Oslo, P. O. Box
1048 Blindern, 0316 Oslo, Norway}
\author{Y.~M.~Galperin}
\affiliation{Department of Physics and Center for Advanced
Materials and Nanotechnology,    University of Oslo, P. O. Box
1048 Blindern, 0316 Oslo, Norway} \affiliation{A. F. Ioffe
Physico-Technical Institute, Polytekhnicheskaya 26, St. Petersburg
194021, Russia}
 \affiliation{Argonne National Laboratory, 9700 S. Cass
Av., Argonne, IL 60439, USA}
\author{T.~H.~Johansen} \email{t.h.johansen@fys.uio.no} \affiliation{Department of Physics and Center
for Advanced Materials and Nanotechnology,   University of Oslo,
P. O. Box 1048 Blindern, 0316 Oslo, Norway}

\date{\today}

\begin{abstract}
The distributions of electrical
current and magnetic field in a thin-film
superconductor ring is calculated
by solving the London equation. The maximum amount of
flux trapped by the hole, the fluxoid saturation number, is obtained by
limiting the current density by the depairing current.
The results are compare it with similar results derived for the bulk case of a
long hollow cylinder
[Nordborg \& Vinokur, \prb {\bf 62}, 12408 (2000)]. In the
limit of small holes our result reduces to the
Pearl solution for an isolated vortex in a thin film.
For large hole radius, the ratio between
saturation numbers in bulk and film superconductors is proportional to
the square root of the hole size.
\end{abstract}


\maketitle

Use of thin film superconductors integrated in nanodevices requires
precise knowledge of the film behavior in the presence of both
external and self induced magnetic fields. Recent experiments have
shown that properly designed arrays of dots and antidots can serve
as effective traps for magnetic flux.\cite{Silhanek05, Grigorenko01,
Berdiyorov06} It has also been shown that patterned superconducting
films allow for the motion of magnetic vortices to be guided over
the film area,\cite{Yurchenko05, Moshchalkov05, Nori03,
Wordenweber97,Moshchalkov98} opening up for a new field often called fluxonics.

The models for flux trapping used in past years did not take into
account the precise geometry of the patterned film samples. Those
conventional models were based on approximations applicable either
to a single vortex in an infinite film, equivalent to having a hole
of zero radius, i.e., a Pearl vortex \cite{Pearl64}, or to an
infinitely long hollow cylinder.\cite{Mkrtchyan72,Nordborg2000} Both
geometries are far from realistic for thin film devices. However
 several recent works considering dynamics of vortices, current and
field distributions in thin-film superconductor with finite hole
size have been
published.\cite{Kirtley2003,Kirtley2003PRL,Babaei2003,Kogan2004,Clem2004}
Yet, the flux saturation number, defined as the maximum number of
flux quanta which can be trapped inside the hole, have so far not
been calculated. Since the saturation number is a key concept for
the vortex-antidot interaction and the overall effect of  film
patterning, we dedicate the present work to this calculation.

In this paper, we examine the problem of trapped flux in a thin-film
superconducting ring by solving the London equations.
The ring geometry allows modelling of a realistic situation with an
isolated anti-dot in a thin-film sample. We consider the case of
zero external magnetic field, and show that the saturation number
for a thin-film ring can differ significantly from that of a hollow
cylinder with the same radius.
It is shown how the difference depends on the ratio between the hole
radius and the thickness of the ring.

\begin{figure}
\centerline{
\includegraphics[width=7cm]{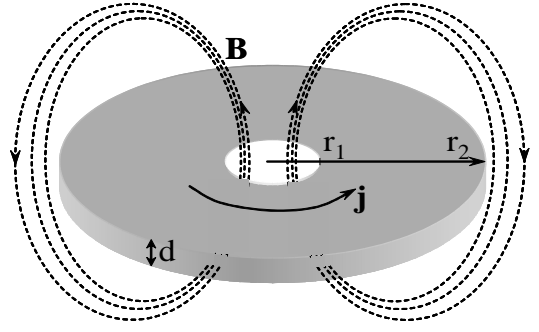}}
\caption{The thin ring geometry. \label{fig:ring}}
\end{figure}

Consider a ring where the outer radius, $r_2$, is much larger than
the inner radius, $r_1$, see Fig.~\ref{fig:ring}.
It is assumed that the film thickness, $d$, is negligible compared
to both the hole radius and the effective penetration length
$\lambda_{\rm eff} = \lambda^2/d$, where $\lambda$ is the London
penetration depth of the superconductor.
We want to determine the amount of flux trapped in the ring in a
remanent state with flux present due to some magnetic prehistory. It
will be assumed that flux pinning elsewhere is absent.

In the superconductor $r_1<r<r_2$, and $-d/2< z < d/2$, the
distributions of current density $\mathbf j$ and induction $\mathbf
B$ are given by the London equation, which in terms of the vector
potential $\mathbf A$, where $\curl \mathbf{A} = \mathbf{B} $, reads
\be \curl (\lambda^2\mu_0 {\mathbf j}+{\mathbf A})=0\, .
\label{eq:LondonRingRot} \ee Due to the symmetry, the current
density and vector potential have in cylindrical coordinates only
azimuthal components, ${\bf j}=(0,j,0)$ and ${\bf A}=(0,A,0)$,
respectively. From the \eq{eq:LondonRingRot} it then follows that
 \be
 \mu_0 \lambda^2 j (r)+A(r)=\frac{\Phi_f}{2\pi r} \, ,
 \label{eq:trapFlux}
 \ee
where flux quantization implies that the constant $\Phi_f$, the
London fluxoid, is a free parameter restricted to an integer number
of the flux quantum,  $ \Phi_f = n \Phi_0$, where $\Phi_0=h/2e$. The
second term on the left-hand-side represents the flux through the
area of radius $r$; $\int_0^{r}{B(r') 2\pi r'\, dr'} = A(r) \, 2 \pi
r$, where $B$ is the induction in the film plane, and is directed
perpendicular to the plane.

\begin{figure}
\centerline{\includegraphics[width=\columnwidth]{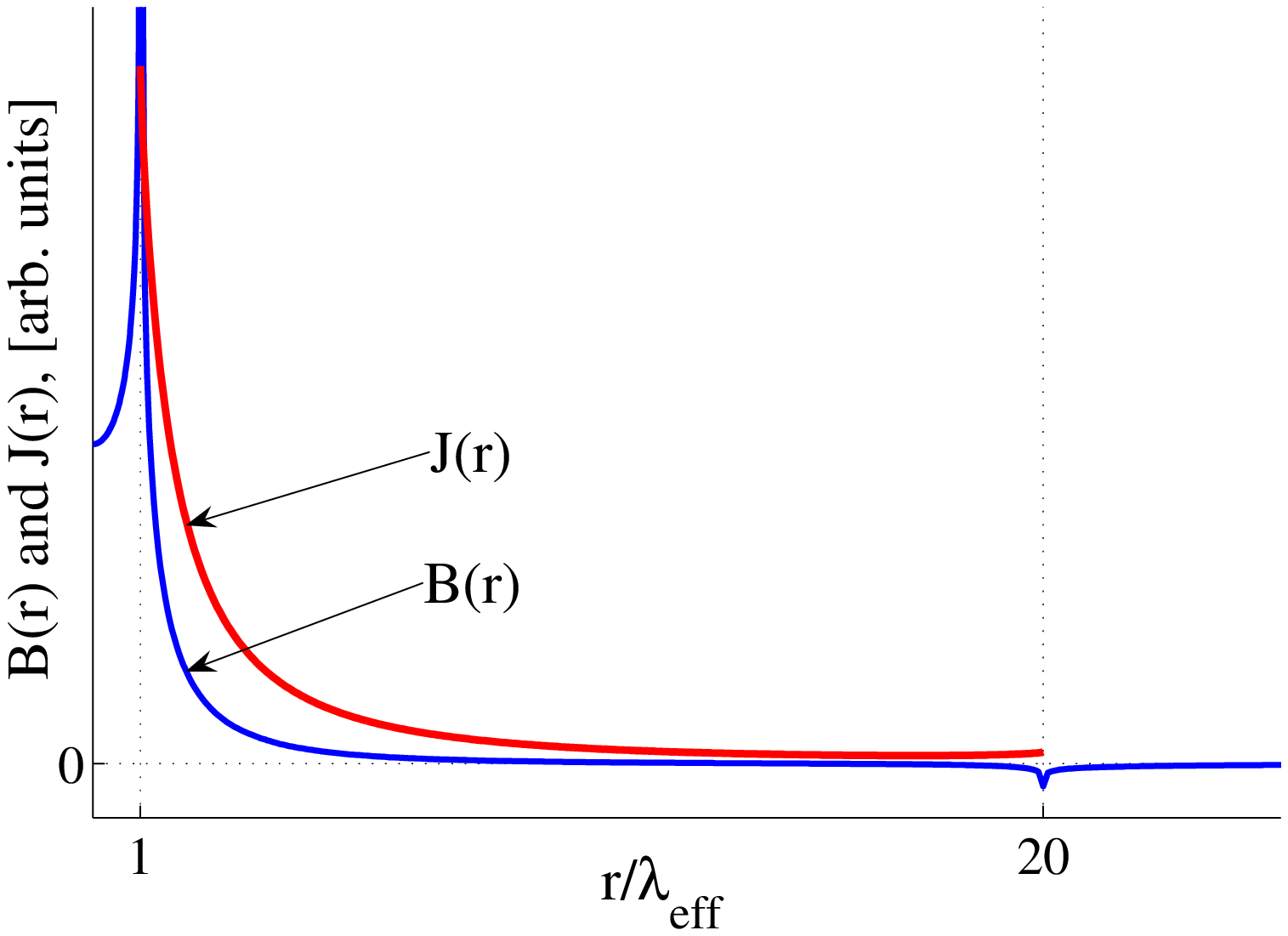}}
\centerline{\includegraphics[width=\columnwidth]{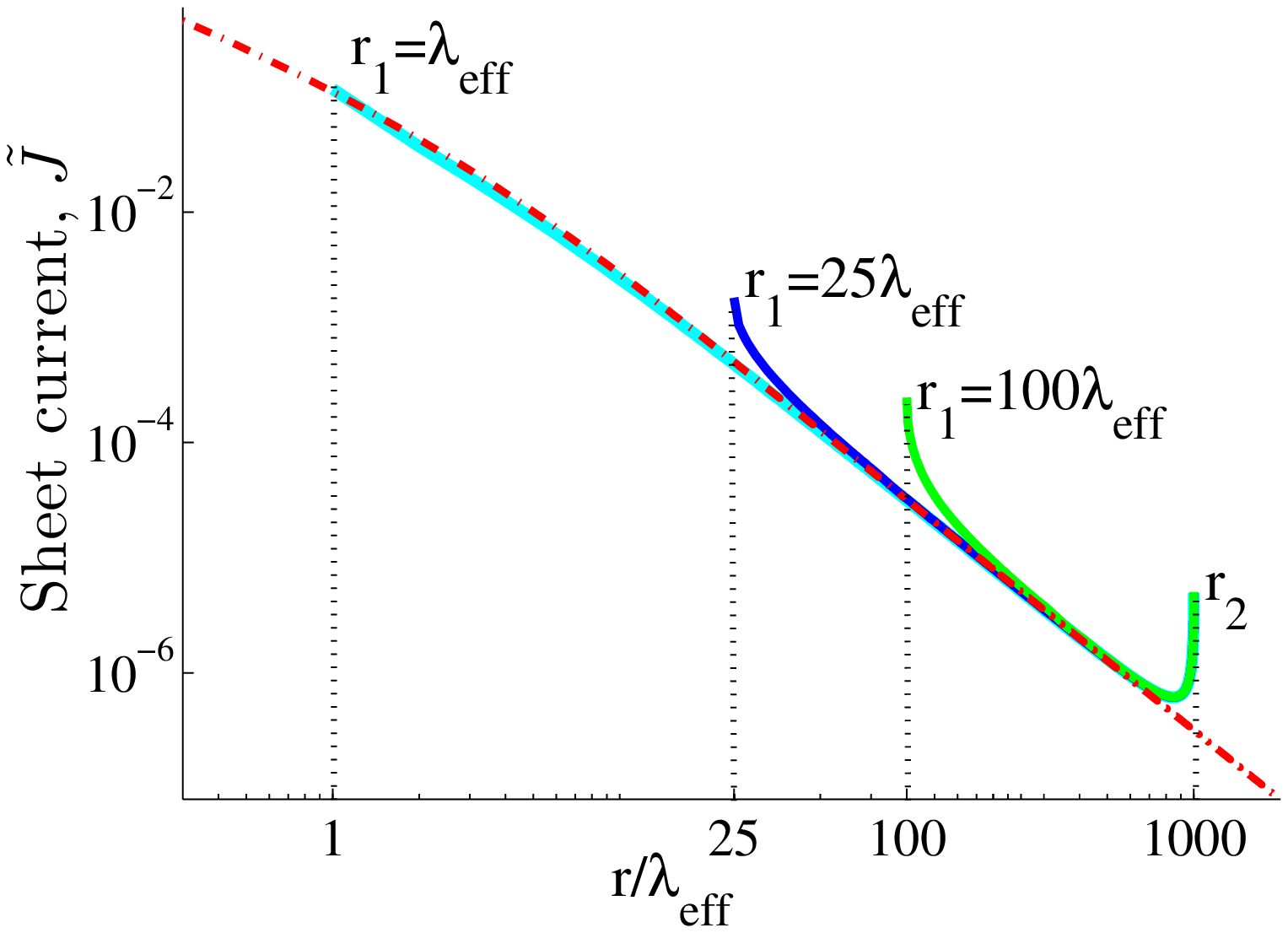}}
\caption{(Upper) Distribution of magnetic field $H$ and current $J$ in the superconducting ring. 
Inner radius of the ring $r_1$ is equal to
$\lambda_\e$ and outer radius $r_2$ is equal to $20\lambda_\e$.
(Lower) Current density distribution in superconducting thin rings
with 3 different hole radii, $r_1$, and the same overall size $r_2$.
For comparison, the graph includes also the Pearl vortex current distribution,
plotted as the dash-dotted line.
\label{fig:a1a25a100}}
\end{figure}

In the absence of an applied field, the vector potential is only due
to the current in the ring, and can be expressed as
\be
{\mathbf
A}=\frac{\mu_0}{4 \pi}\int{\frac{\mathbf{j}} {|\mathbf{r}-{\mathbf
r}'|}}  d^3r'   \, .
\label{eq:Atojvec}
\ee
For the film geometry
one can neglected variations of the current across the the sample
thickness and average \eq{eq:Atojvec} over $z$. Introducing the
sheet current
 $J(r)=\int_{-d/2}^{d/2} j(r)dz$,  together with the dimensionless variables
 $\tilde{J}(r)=(\mu_0 \lambda_\e^2/\Phi_f)J(r)$
and $\tilde{r}=r/\lambda_\e$ the \eq{eq:trapFlux} becomes,

\begin{eqnarray}
&& \tilde{J}(\tilde{r}) +
\frac{1}{4\pi}\int_{\tilde{r_1}}^{\tilde{r_2}}\! \! \int_0^{2\pi}\!
\!
{\frac{\tilde{J}(\tilde{r'})\cos\theta}{\sqrt{(\tilde{r}/\tilde{r}')^2+
1-2(\tilde{r}/\tilde{r}')\cos\theta}}\,
  d\theta d\tilde{r}'}
\nonumber \\ && \quad \quad =1/2\pi \tilde{r}\, .
\label{eq:dimension}
\end{eqnarray}
This Fredholm integral equation of the second kind was solved
numerically by converting it into a set of linear equations
  corresponding to discrete values of the radial coordinate,
\begin{eqnarray}
\tilde{J}_i&+&\frac{1}{4 \pi}\sum_{ij}Q_{ij}\tilde{J}_j=\frac{1}{2\pi \tilde{r}_i}  , \\
\label{eq:linearset}
Q_{ij}&\equiv&\int_0^{2 \pi}{\frac{\cos\theta}{\sqrt{(\tilde{r}_i/\tilde{r}_j)^2+1-2(\tilde{r}_i/\tilde{r}_j)\cos\theta}}d\theta} \, .
\label{eq:kernel}
\end{eqnarray}
Results of such calculations are presented in
Fig.~\ref{fig:a1a25a100}. The upper panel shows the field and
current distribution for the case where $r_1 = \lambda_\e$, $r_2 =
20 \lambda_\e$. The field is mostly concentrated in the hole, where it has
a pronounced peak at $r_1$, while a small peak from the return field
is seen near $ r_2$. The shielding current flows predominantly in
the vicinity of the hole, reaching a maximum at the edge.

Plotted in the lower panel is the normalized current distribution
for holes of different sizes ranging from $r_1/\lambda_\e$ = 1 to
100, with all rings having $r_2/\lambda_\e = 1000$. For comparison,
the figure includes also
 the current distribution around a single Pearl
vortex in an infinite film,~\cite{Pearl64} \be
\tilde{J}_{\text{Pearl}}(\tilde{r})=\left[S_1(\tilde{r}/2)-
K_1(\tilde{r}/2)-2/\pi\right]/8\, .
\label{eq:currentPearl}
\ee
Here $S_1$ is the first order Struve function and $K_1$ the first order
modified Bessel function of the second kind. For the small hole
case, $r_1 = \lambda_\e $, our numerical result is very close to the
Pearl solution over the whole ring area except near the outer edge,
where the sheet current has an upturn. The curves for the two larger
hole radii also follow \eq{eq:currentPearl} for intermediate $r$,
but are seen to have an additional upturn at the inner edge. The
current enhancement near $r_1$  is found to increase with the hole
size, whereas near the outer edge the behavior is marginally
influenced by the hole. Similar results for superconducting rings
were reported previously by Brandt and Clem\cite{Clem2004}.

Our main interest lies in finding the fluxoid saturation value as
function of the size of the hole in a thin film.
It follows from \eqs{eq:trapFlux}{eq:Atojvec}
 that the magnitude of the sheet current depends linearly on
$\Phi_f$. Thus, the maximum $\Phi_f$ corresponds to having a current
at the inner edge, $J(r_1)$, equal to the maximum current supported by the
superconductor. We take this maximum value  to be the depairing
current,~\cite{UC}
\be
j_{dp}=\frac{\Phi_0 }{3\sqrt{3}\pi \mu_0 \lambda_\e \xi}\,.
\label{eq:uncoupling}
\ee
Thus, the fluxoid saturation number, $n^{\fl}$, satisfies,
\be
\frac{n^{\fl} \Phi_0}{j_{dp}d} = \frac{\Phi_f}{J(r_1)} \, ,
\ee
which gives
\be
n^{\fl}= \frac{\lambda_\e}{3\sqrt{3}\pi \xi \tilde{J}(\tilde{r}_1)}\,
 \label{eq:nquanta} \, .
\ee

Shown in \f{fig:nquanta} is the fluxoid saturation number as a function of the hole size.
The following material parameters were here used, $\xi=3$~nm, $\lambda=150$~nm, $d=100$~nm,
corresponding to YBa$_2$Cu$_3$O$_x$ at low temperatures.
The stepwise increase of $n^{\fl}$ versus $r_1$ seen in the main plot is essentially linear,
but with an additional weak upward curvature. The inset shows that this behavior continues
also as for larger holes,  $r_1 > \lambda$.


For small holes 
we found that the current
near the inner edge behaves essentially according to \eq{eq:currentPearl},
which in this limit is well approximated
by $\tilde{J}(\tilde{r}) = 1/2 \pi \tilde{r}$.
Thus, for small holes the saturation number is given by,
\be
n^{\fl}=\frac{2\,r_1}{3\sqrt{3}\, \xi} \ \ , \ \
r_1\ll \lambda_\e \, .
\label{eq:ll1}
\ee
For larger holes where $r_1 \gg \lambda_\e$, the current upturn near the edge
prevents us from using the Pearl solution directly.
Instead, we make a conjecture that the sheet current distribution is
described by
\be
\tilde{J}(\tilde{r})=\frac{1}{\pi
  \tilde{r}\sqrt{\tilde{r}^2-\tilde{r}_1^2}}\,.
\label{eq:Iapprox}
\ee
Firstly, this analytical form fits excellently when compared to the current enhancement
calculated numerically.
Secondly, this form has the same asymptotic behavior as the Pearl solution,
 $\tilde{J}(\tilde{r}) = 1/\pi \tilde{r}^2$ for $\tilde{r} \gg 1$, consistent with the close
agreement seen in Fig. \ref{fig:a1a25a100} over a wide intermediate range of $r$.
Finally, the \eq{eq:Iapprox} has the
diverging factor  $1/\sqrt{r-r_1}$, commonly present in the edge behavior of
thin superconductors in the Meissner state.\cite{Divergence}
As usual, the divergence is cut off at a distance $\lambda_\e$ from the edge,
giving for the ring geometry a maximum sheet current of
$\tilde{J}(\tilde{r}_1)=1/(\sqrt{2}\pi\tilde{r_1}^{3/2})$.
It then follows from \eq{eq:nquanta} that for large holes one has,
\be
n^{\fl}= \frac{\sqrt{2}\, r_1^{3/2}}{3\sqrt{3} \, \xi
\lambda_\e^{1/2}} \ , \ \ \ \  r_1\gg\lambda_\e\, .
\label{eq:Nfilm}
\ee

\begin{figure}
\centerline{
\includegraphics[width=\columnwidth]{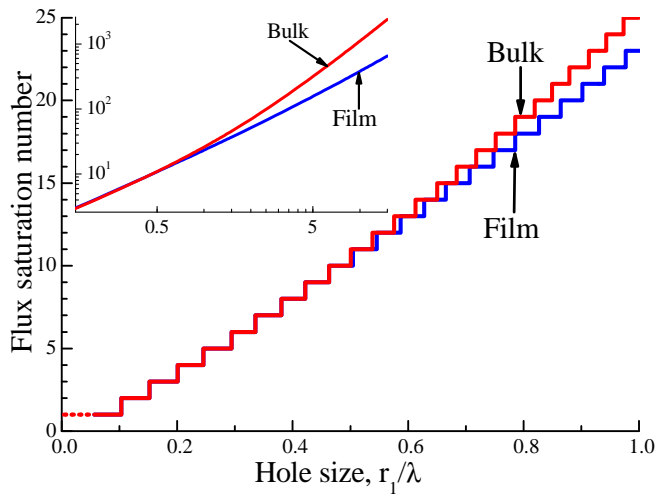}
}
\caption{Trapped flux inside the hole of thin superconducting film (blue line)
and inside the infinite cavity in bulk superconductor (red line) depending on the size of the hole/cavity.
\label{fig:nquanta}}
\end{figure}

It is of interest to compare these results with the fluxoid saturation
number associated with a hole in a bulk superconductor. The case of
a circular hole through an infinite superconductor was discussed in
Ref.~\onlinecite{Nordborg2000}, using that for this geometry
with $r_1$ being the hole radius, the induction is given by
\be
B(r)= n \Phi_0 C K_0(r/\lambda) \ \ , \ \  r \geq r_1 \, ,
\label{eq:BNordberg}\\
\ee
where  $1/C = \pi r_1 \lambda [2K_1(r_1/\lambda)+(r_1/\lambda) K_0(r_1/\lambda)]$,
and $K_0$, $K_1$ are the zeroth-order and first-order modified Bessel functions.
The current density around the hole here obtained simply by,
$\mu_0 j(r)=B'(r)$, and with $j(r_1) = j_{dp} $,
 the maximum number of trapped flux quanta becomes,
\be
n_{\bl}=
\frac{2}{3\sqrt{3}}\frac{r_1}{\xi}\left[1+\frac{r_1}{2\lambda}\frac{K_0(r_1/\lambda)}{K_1(r_1/\lambda)}\right]\, .
\label{eq:nbulk}
\ee

The bulk saturation number is also plotted in Fig. \ref{fig:nquanta}, where one sees the same
tendency as for the thin film case, namely that a larger hole size allows more flux to be trapped.
For small $r_1$ the two curves are actually overlapping. This is because
 in \eq{eq:trapFlux}, which is valid for any cylindrical geometry,
the vector potential term becomes negligible for small  $r$,
giving $\mu_0\lambda^2 j(r)2\pi r=\Phi_f$ as the asymptotic behavior
independent of sample thickness. Thus, the linear hole size dependence in
\eq{eq:ll1}
holds also for the bulk case.

As $r_1$ increases the curves deviate,
showing that thin film superconductors will not trap as much flux
as a bulk sample of the same material.
For $r_1/\lambda\gg 1$, the
Bessel functions in \eq{eq:nbulk} can be simplified,
and one gets asymptotically a quadratic hole size dependence,
\be
n_{\bl} =  \frac{r_1^2 }{3\sqrt{3}\xi \lambda}  \ \ \ , \ \ \
r_1 \gg \lambda  \, .
\label{eq:Ngg1}
\ee
 For large holes, the ratio in flux trapping capability is simply
\be
\frac{n_{\bl}}{n^{\fl}}=\sqrt{\frac{r_1}{2d}}\, .
\label{eq:nbnf}
\ee
For typical patterned films with hole size,\cite{Silhanek05} $r_1 = 500 - 600$ nm,
this ratio is close to 2.

 Note  that when the current density approaches
the depairing value, the superconducting order parameter will be
reduced. In particular, this will occur most easily in the vicinity of the hole.
This means that one should consider the saturation numbers derived
in this work as upper limiting values for the number of flux quanta
possibly trapped by the superconductor.

In summary, we have derived the fluxoid saturation number for thin
superconducting films containing a circular hole of finite radius. For
 small holes, $r_1\ll \lambda_\e$, the result is approximately
equal to the flux saturation number for bulk superconductors
with a cylindric cavity. For
large holes, $r_1 \gg \lambda_\e$, the saturation number in
thin films is less than in bulks by a factor $\sqrt{r_1/2d}$.
The present results, obtained by assuming the sample size to be much
larger than the hole,
should apply to superconducting films patterned with
a single antidot. Even more, knowing the fluxoid saturation number,
one may predict the interaction between vortices and antidot arrays
in various experimental configurations.

\acknowledgements This work was partly supported by the Norwegian
Research  Council and by the U. S. Department of Energy
Office of Science contract No. DE-AC02-06CH11357, and by the
Russian Foundation for Basic Research grant 06-02-16002. We are
thankful to O. Fefelov and V. Vinokur for helpful discussions, and
to J. R. Clem for valuable comments.

\end{document}